\RequirePackage{fix-cm}
\documentclass{svjour3}                     
\smartqed  
\usepackage{graphicx}
\journalname{Microfluidics and Nanofluidics}
\usepackage{amsmath}
\usepackage{lipsum}
\usepackage{filecontents}
\usepackage[usenames,dvipsnames,svgnames,table]{xcolor}
\usepackage{graphicx}
\usepackage[caption=false]{subfig}
\usepackage{comment}
\usepackage[utf8]{inputenc}
\usepackage[english]{babel}
\usepackage[T1]{fontenc}
\usepackage{chngcntr}
\usepackage{appendix}
\usepackage{ifpdf}
\ifpdf
  \DeclareGraphicsExtensions{.pdf,.png,.jpg}
\else
  \DeclareGraphicsExtensions{.eps}
\fi
\usepackage[scale=0.77]{geometry}
\usepackage{color,soul}
\usepackage[framemethod=tikz]{mdframed}
\usepackage{import}
\usepackage{float}
\usepackage{layout}
\usepackage{bm}

\usepackage{hyperref}

\setstcolor{red}

\begin{document}

\title{Experimental and theoretical investigation of a low-Reynolds-number flow through deformable shallow microchannels with ultra-low height-to-width aspect ratios
}


\author{Aryan Mehboudi        \and
        Junghoon Yeom 
}


\institute{A. Mehboudi \at
              Mechanical Engineering Department, Michigan State University, East Lansing, Michigan, US \\
           \and
           J. Yeom \at
              Mechanical Engineering Department, Michigan State University, East Lansing, Michigan, US \\
              Tel.: +1 517-432-9132\\
              Fax: +1 517-353-1750\\
							\email{jyeom@egr.msu.edu}           
}

\date{(Submitted to Microfluidics and Nanofluidics: March 25, 2018. Today is \today.)}

\maketitle

\begin{abstract}

The emerging field of deformable microfluidics widely employed in the Lab-on-a-Chip and MEMS communities offers an opportunity to study
a relatively under-examined physics. 
The main objective of this work is to provide a deeper insight into the underlying coupled fluid-solid 
interactions 
of a low-Reynolds-number, {\it i.e.} Re$\sim O(10^{-2}$\textemdash$10^{+1})$, fluid flow through a shallow deformable microchannel with ultra-low height-to-width-ratios, {\it i.e.} $O(10^{-3})$. 
The fabricated deformable microchannels of several microns in height and few millimeters in width, whose aspect ratio is about two orders of magnitude smaller than that of the previous reports, allow us to investigate the fluid flow characteristics spanning a variety of distinct regimes from small wall deflections, where the deformable microchannel resembles its corresponding rigid one, to wall deflections much larger than the original height, where the height-independent characteristic behavior emerges. 
The effects of the microchannel geometry, membrane properties, and pressure difference across the channel are represented by a lumped variable called flexibility parameter.
Under the same pressure drop across {different} channels, any difference in their geometries is reflected into the flexibility parameter of the channels, which can potentially cause the devices to operate under distinct regimes of the fluid-solid characteristics.
For a fabricated microchannel with given membrane properties and channel geometry, on the other hand, 
a sufficiently large change in the applied pressure difference 
can alter the flow-structure behavior from one characteristic regime to another.
By appropriately introducing the flexibility parameter and the dimensionless volumetric flow rate, a master curve is found for the fluid flow through any long and shallow deformable microchannel.
A criterion is also suggested for determining whether the coupled or decoupled fluid-solid mechanics should be considered.

\end{abstract}

\newpage
\section[Introduction]{Introduction}
\label{Introduction} 

While the rapidly-growing microfluidics technology has already permeated through many aspects of the molecular and biological sciences, the emerging field of deformable microfluidics has been attracting an increasing attention during the last decade, 
enabling a variety of different applications such as 
developing fluidic circuits \cite{leslie_frequency-specific_2009,lam_microfluidic_2006}, 
tunable optofluidic devices \cite{song_elastomer_2012}, 
size-tunable droplet generation \cite{abate_valve-based_2009,raj_droplet_2016}, 
particle/cell separation 
\cite{huang_tunable_2009,gerhardt_chromatographic_2011,beattie_clog-free_2014,huang_clogging-free_2014},
etc.
Shallow microchannels with a low height-to-width ratio, on the other hand, are important for various bio-applications such as the label-free cell detection \cite{javanmard_targeted_2007}, immunoassays \cite{wang_ultrasensitive_2013,smith_research_2014}, cell culturing \cite{jang_selective_2012}, etc., where the underlying detection mechanism is based on the biospecies interaction with the innermost surface of the microchannel.
Under these situations, the performance or detection limit can be improved by increasing the probability of interactions of the target species and the wall surfaces through decreasing the channel height and/or increasing the channel width.

Even though there are several experimental reports on the fluid-solid mechanics of a fluid flow within a deformable shallow microchannel \cite{cheung__2012,ozsun_non-invasive_2013,seker_nonlinear_2009,hardy_deformation_2009,kang_pressure-driven_2014,neelamegam_experimental_2015,raj_hydrodynamics_2017}, measuring the velocity and pressure throughout the channel is still a serious challenge.
The two-/three-dimensional mathematical models have been, therefore, used to better understand the pertinent physics \cite{chakraborty_fluid-structure_2012,shidhore_static_2018}.
For the mathematical modeling with the mutual fluid-solid interactions taken into account, the pressure, as a property from the fluid mechanics domain, affects the membrane deflection, as a parameter belonging to the solid mechanics domain.
On the other hand, any change in the microchannel geometry can alter the pressure and the velocity fields throughout the microchannel.
Hence, the solid and fluid aspects of this problem can be potentially highly coupled, which makes it hard, if not impossible, to present an analytical solution for the two-/three-dimensional models.
The governing equations are, therefore, solved numerically, which can be very time consuming \cite{shidhore_static_2018}, particularly for the shallow microchannels with the ultra-low height-to-width-ratios because the grid size needs to be sufficiently small to capture the variations of the parameters across the channel's small height as well as its large width and length.

One-dimensional modeling is considered as a fast and convenient tool for studying and designing the deformable microchannels.
Gervais et al. \cite{gervais_flow-induced_2006} proposed a mathematical model for the rectangular microchannels with a thick deformable ceiling by means of the scale analysis.
They experimentally corroborated the proposed model with a fitting parameter that led to the acceptable agreement between the experiments and the modeling.
The fitting parameter is dependent on the channel geometry as well as the physical properties of the membrane and working fluid. 
Therefore, the model with the fitting parameters requires experimental results and cannot be widely used \cite{hardy_deformation_2009,christov_flow_2018}.
Relying on several assumptions including 
the parabolic shape for the deflection profile of the thin membrane wall, 
Raj and Sen \cite{raj_flow-induced_2016} developed a one-dimensional model expressing the relationship between the volumetric flow rate and the pressure difference across the microchannel, without the need of the fitting parameters.

Christov {\it et.al.} \cite{christov_flow_2018} have more recently 
developed a one-dimensional coupled fluid-solid mechanics model using the lubrication theory.
The main idea behind the model is to assume each infinitesimal slice of the membrane (normal to the flow-wise direction) independently deflects under the corresponding local pressure of the working fluid within the microchannel.
From the solid-mechanics point of view, an infinitesimal slice of the membrane is replaced by  
a wide beam, 
the deflection profile of which is determined as a function of the local pressure.
This assumption necessitates the microchannel length to be much larger than its width, 
which is also a basic assumption for the lubrication theory.
From the fluid-mechanics point of view, the lubrication theory suggests the Poiseuille flow to be locally valid for 
the equivalent channel with the same cross sectional area dictated by the deflection profile.
This approximation provides a first-order ordinary differential equation, and from its solution, a relationship is extracted between the local pressure and the volumetric flow rate.
Even though this framework was initially applied to the thin-plate-bending-dominated ceiling deformation, Shidhore and Christov \cite{shidhore_static_2018} have implemented the same strategy to other elastic deformation regimes of thick-plate-bending and stretching. 

Christov {\it et.al.} \cite{christov_flow_2018} demonstrated the volumetric flow rate can be written as shown in Eq.~\ref{Eq-General-Christov}, 
where $Q$ is the flow rate, $\Delta p$ is the pressure drop, $W$ is the channel width, $H_0$ is the
undeformed channel height, $L$ is the channel length, $\mu$ refers to the fluid dynamic viscosity, and $D = {Et^3}/{12(1-\nu^2)}$, in which $t$, $E$, and $\nu$ refer to the membrane thickness, modulus of elasticity, and  Poisson's ratio, respectively:
\begin{equation}
\begin{aligned}
Q=\frac{\Delta pWH^3_0}{12\mu L}\Bigg[1 &+ 
\frac{1}{480}\frac{W^4}{DH_0}\Delta p \\ 
&+\frac{1}{362,880}\bigg(\frac{W^4}{DH_0}\bigg)^2\Delta p^2\\ 
&+\frac{1}{664,215,552}\bigg(\frac{W^4}{DH_0}\bigg)^3\Delta p^3\Bigg]  
\end{aligned}
\label{Eq-General-Christov}
\end{equation}
By keeping only the leading-order elasticity contribution in Eq.~\ref{Eq-General-Christov}, Christov {\it et.al.} \cite{christov_flow_2018} have summarized the key result for the bending-dominated deformation as: 
$Q\approx \frac{\Delta pWH^3_0}{12\mu L}\big[1+\frac{3}{160}(\frac{W}{t})^3(\frac{W}{H_0})(\frac{\Delta p}{E})\big]$, 
assuming an incompressible material, {\it i.e.} the Poisson ratio $\nu=1/2$.
However, they have also pointed out in their paper that for a sufficiently large value of $\Delta p{W^4}/{DH_0}$, the higher-order
terms become important in Eq.~\ref{Eq-General-Christov}.
To the best of our knowledge there is no literature studying the low-Reynolds-number flows under the large regimes of $\Delta p{W^4}/{DH_0}$ perhaps due to the lack of the fabrication method to create the ultra-low height-to-width-ratio deformable microchannels.
Most of the deformable microchannels like those investigated by Shidhore and Christov \cite{shidhore_static_2018} show the relatively high Reynolds numbers and small values of $\Delta p{W^4}/{DH_0}$.
The large regimes of this parameter and its effects on the fluid-solid mechanics behavioral characteristics of the low-Reynolds-number fluid flows still remain unexplored.

In this work, we experimentally and mathematically investigate the {\it volumetric flow rate\textemdash pressure difference} relationship for the shallow deformable microchannels with ultra-low height-to-width-ratios.
The microchannels of few millimeters in width and few microns in height are fabricated by modifying our previously reported fabrication protocol \cite{mehboudi_two-step_2018} to enable the membrane to freely deform, while the bonding reinforcement allows the high pressure differences to be applied.
This fabrication method enables us to uncover new fluid-solid behavioral characteristics for the low-Reynolds-number flow, {\it i.e.} Re$\sim O(10^{-2}$\textemdash$10^{+1})$.

Our experimental results combined with the scale analysis of the model proposed by Christov {\it et.al.} \cite{christov_flow_2018} 
reveal that various distinct fluid-structural characteristics emerge under different regimes of the flexibility parameter.
The flexibility parameter is a lumped variable in this work, reflecting the effects of different parameters such as the microchannel dimensions, membrane properties, and pressure difference across the channel.
We show that introducing the flexibility parameter along with an appropriate form of dimensionless volumetric flow rate leads to a master curve, which is valid for any arbitrary shallow and long deformable microchannel.
For the sufficiently small flexibility parameter, the characteristics match {with} those of the fluid flow through the rigid microchannels, while the {\it height-independent} characteristics emerge for the sufficiently large flexibility parameter. 
It is observed that the differences between the microchannel geometries 
result in the different flexibility parameters under the same pressure difference across the channels, 
while a significantly large change in the pressure difference across a fabricated microchannel with given geometry and membrane properties can alter the flexibility parameter scale from one characteristic regime to another.
It is shown that the microchannel width plays an important role in the 
fluid-solid characteristics;
a relatively small difference in the microchannels widths ($\sim 60\%$) can result in a one-order-of-magnitude difference in the flexibility parameters, which can potentially cause two distinct fluid-solid behavioral characteristics. 
Our further analysis provides a threshold for the flexibility parameter, by which the coupled and decoupled fluid-solid mechanics regimes are separated.
Below this threshold, the pressure varies linearly through the microchannel, while the assumption of linear pressure variation becomes noticeably erroneous for the large flexibility parameters, since the coupled fluid-solid interactions enforce a non-linear pressure distribution within the channel.

The rest of this paper is organized as follows.
In section \ref{Modeling} the mathematical aspect of this investigation is described.
The fabrication and experimental setup are illustrated in section \ref{Experiments}.
Results and discussion are presented in section \ref{Results and discussion}.
We conclude with a brief summary in section \ref{Conclusion}.

\section[Modeling]{Modeling}
\label{Modeling}

\subsection[Analytical one-dimensional coupled fluid-solid mechanics model]{Analytical one-dimensional coupled fluid-solid mechanics model}
\label{Analytical one-dimensional coupled fluid-solid mechanics model}

In this section, we revisit the one-dimensional coupled fluid-solid mechanics model developed by Christov {\it et. al.} \cite{christov_flow_2018} for the fluid flow through the shallow microchannels.
The governing equations and parameters are, however, rearranged to provide a better insight into the underlying physics.

\begin{figure}[t!]
	\centering
	\subfloat{
	\includegraphics[width=1\linewidth]{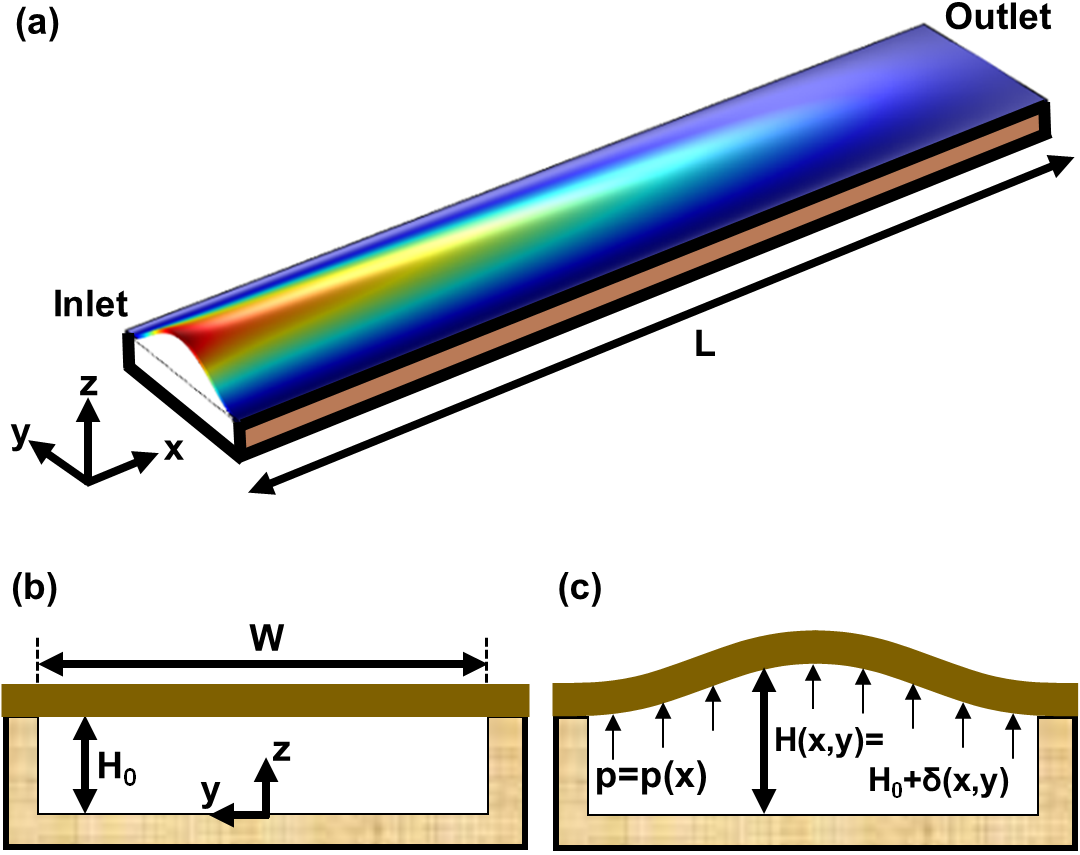}
	}
	\caption{(a) Schematic representation of a fluid flow through a microchannel with the flexible ceiling together with the cross-section view of the channel (b) without and (c) with applying the pressure difference across the channel.
	}
	\label{Fig-Straight-Schem}
\end{figure}

The schematic presentation of a fluid flow through a straight microchannel with a deformable ceiling is shown in Fig.~\ref{Fig-Straight-Schem}.
The model developed by Christov {\it et. al.} \cite{christov_flow_2018} is based on the lubrication theory, {\it i.e.} 
${H\textsubscript{0}}\ll {W}\ll {L}$, 
where 
${H\textsubscript{0}}$, ${W}$, and ${L}$
 refer to the microchannel original height with no membrane deformation, width, and length, respectively. 
From the various types of the elastic deformation studied by Shidhore and Christov \cite{shidhore_static_2018} such as thin-membrane-bending, thick-plate-bending, stretching, etc., the {\it thin-plate-bending framework} is utilized in this work, since it suits the configuration of the fabricated microchannels with a deformable thin PET 
(polyethylene terephthalate) 
film.
This framework requires 
$\delta\text{\textsubscript{max}}\ll {t}\ll {W}$, 
where $\delta\text{\textsubscript{max}}$ is the maximum displacement of the membrane and ${t}$ denotes the membrane thickness. 

It should be noted that 
despite the three-dimensional nature of the membrane deformation and the velocity field in a general fluid flow through deformable channels, 
because of the assumptions made in~\cite{christov_flow_2018}~such as~${H\textsubscript{0}}\ll {W}\ll {L}$,~the small components of velocity in y and z directions (Fig.~\ref{Fig-Straight-Schem}) do not appear in the leading-order equations using the perturbation technique.
The pressure and the flow-wise component of velocity are the only variables associated with the fluid-mechanics aspect of the problem appearing in the leading-order equations of the perturbation method.
Regarding the solid-mechanics domain, because of the similar assumptions, the ceiling deformation's dependence on the upstream and downstream membrane slices does not appear in the leading-order equations of the perturbation method, leading to the one-dimensional Euler--Bernoulli beam theory through correlating the membrane deformation in each slice with the local pressure load.
Because of the aforementioned reasons, we still refer to the model developed by Christov~{\it et. al.} \cite{christov_flow_2018}~and its extension studied by Shidhore and Christov~\cite{shidhore_static_2018}~as one-dimensional models.

The volumetric flow rate--pressure difference relation obtained by Christov {\it et. al.} \cite{christov_flow_2018}, presented in Eq.~\ref{Eq-General-Christov}, can be rearranged as shown in Eq.~\ref{Eq-sol-gen-Non-dim-straight-Q-flex-over-Q-rig-analytical}, in which $Q_\text{Deformable}$ and $Q_\text{Rigid}$ respectively refer to the volumetric flow rate through the deformable microchannel and that through the corresponding rigid microchannel
\begin{equation}
\begin{aligned}
Q^\ast=\frac{Q_\text{Deformable}}{Q_\text{Rigid}}&=1+c_1\chi+c_2{\chi}^2+c_3{\chi}^3\\
c_1&=\frac{4}{5}=0.8\\
c_2&=\frac{128}{315}\approx 0.4063\\
c_3&=\frac{256}{3003}\approx 0.0852
\end{aligned}
\label{Eq-sol-gen-Non-dim-straight-Q-flex-over-Q-rig-analytical}
\end{equation}
where we define the dimensionless parameter $\chi={\Delta pW^4}/{384DH_0}$, namely a {\it flexibility parameter} hereafter.
In addition,
$D = {Et^3}/{12(1-\nu^2)}$, 
in which $E$, and $\nu$ refer to the membrane modulus of elasticity and  Poisson's ratio, respectively.
The volumetric flow rate through the rigid microchannel can be written as $Q_\text{Rigid}=\Delta pWH^3_0/12\mu L$, 
where $\mu$ is the fluid dynamic viscosity.
By defining the flexibility parameter and the dimensionless volumetric flow rate in this manner, we aim at directly observing the effects of 
various dimensional or operational parameters  
on how different the flow characteristics
of a deformable microchannel and those of its theoretically rigid counterpart 
are. 
Such differences between the rigid and deformable microchannels are reflected into the magnitude of $Q^\ast$. 
The minimum value of $Q^\ast$ is 1, which corresponds to a theoretically rigid microchannel. 
A configuration with $Q^\ast\approx 1$ shows that the channel deformability effects are negligible and flow characteristics are close to those of the flow through the rigid counterpart, while an increase in $Q^\ast$ implies a more significant deviation from the rigid channel characteristics.

The mean and maximum dimensionless displacements of the membrane are expressed in Eq.~\ref{Eq-deflection-mean-straight}:
\begin{equation}
\begin{aligned}
\frac{<\delta({x^\ast})>}{H_0}&=\frac{8}{15}\chi~p^\ast(x^\ast)\\
\frac{\delta_\text{max}({x^\ast})}{H_0}&=\chi~p^\ast(x^\ast)
\end{aligned}
\label{Eq-deflection-mean-straight}
\end{equation}
where $p^\ast=p/\Delta p$ shows the dimensionless pressure within the channel, and $x^\ast=x/L$ is the dimensionless distance from the inlet.
The parameters $<\delta({x^\ast})>$ and $\delta_\text{max}({x^\ast})$ denote the mean and maximum membrane displacement at ${x^\ast}$, respectively.

\subsection[Composite membranes]{Composite membranes} 
\label{Composite membranes}

\begin{figure}[t!]
	\centering
	\subfloat{
	\includegraphics[width=1\linewidth]{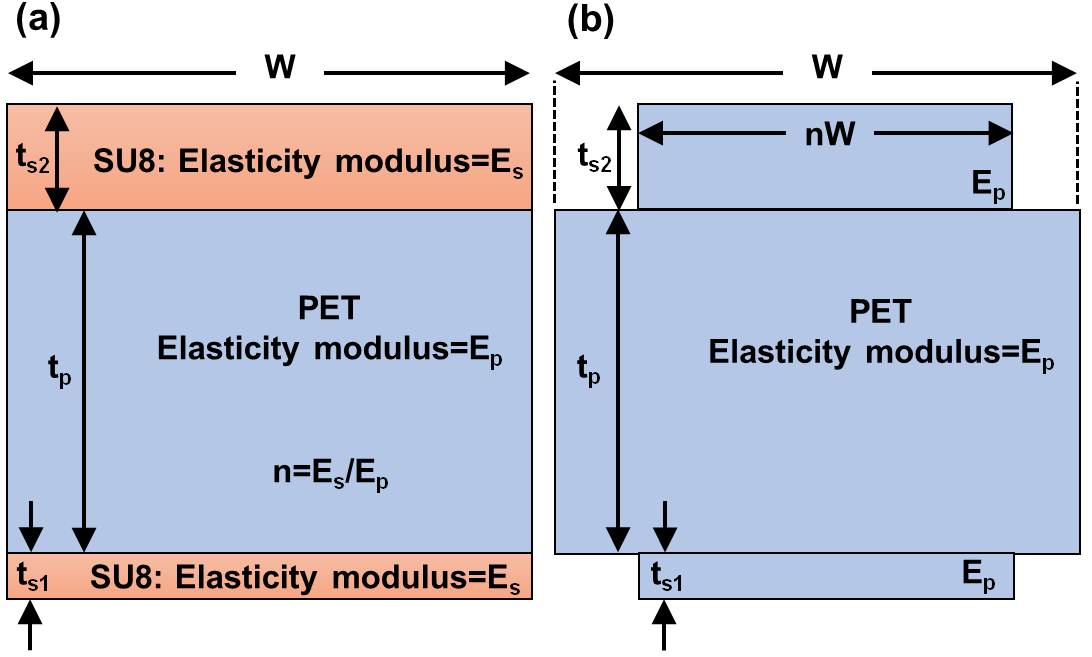}
	}
	\caption{(a) Schematic diagrams of the cross section of the composite membrane consisting of the PET film with two SU8 adhesive layers together with the (b) transformed section for stress analysis.
	}
	\label{Fig-Schem-transformed-section}
\end{figure}

Because of the adhesive bonding technique used in this work, the structural effects of the adhesive layers need to be taken into account.
As will be described in section~\ref{Fabrication}, the microchannel ceiling consists of a PET 
film coated with SU8 on both sides, forming a three-layer membrane; see Fig~\ref{Fig-Schem-transformed-section} (a). 
By using the transformed-section method \cite{gere_mechanics_2004}, the flexural rigidity $F=EI$, where $I$ denotes the moment of inertia, is obtained for the transformed section; see Fig~\ref{Fig-Schem-transformed-section} (b).
The equivalent membrane thickness, $\tilde{t}$, is calculated from $\tilde{t}=\sqrt[3]{12F/EW}$, and replace the membrane thickness, $t$, to evaluate the flexibility parameter, $\chi$, in Eq.~\ref{Eq-sol-gen-Non-dim-straight-Q-flex-over-Q-rig-analytical}.
The composite membrane is then considered as a single-layer film with the equivalent thickness of $\tilde{t}$, {\it i.e.} $t$ is replaced by $\tilde{t}$, and the elasticity modulus of PET, {\it i.e.} $E$ is replaced by $E_\text{PET}$.

\subsection[Linear pressure variation assumption]{Linear pressure variation assumption} 
\label{Linear pressure variation assumption}

The membrane displacement is dictated by the pressure variation within the microchannel.
On the other hand, the membrane displacement changes the cross sectional area and in turn hydrodynamic resistance, which influences the pressure variation through the channel.
Hence, the fluid and solid aspects of this problem are potentially highly coupled.
However, the fluid and solid aspects can be studied in a decoupled fashion when the membrane deformation is sufficiently small compared to the microchannel original height.
This decoupled state arises with a small pressure difference, deep and/or narrow channel, or thick/stiff membrane.
Under these situations, the pressure varies almost linearly within the microchannel.
By substituting 
the linear pressure variation of $p^\ast(x^\ast)=1-x^\ast$ in Eq.~\ref{Eq-deflection-mean-straight}, we have
$\frac{<\delta {(x^\ast)}>}{H_0}=\frac{8}{15}\chi~(1-x^\ast)$.
The average of the membrane displacement over the whole channel can be obtained through 
$\int^{x^\ast=1}_{x^\ast=0}<\delta {(x^\ast)}>dx^\ast=\frac{4}{15}\chi H_0$.
If the deformable microchannel is replaced with a rigid rectangular microchannel with the height of $H_0+\frac{4}{15}\chi H_0$, the volumetric flow rate can be expressed as in Eq.~\ref{Eq-vfr-lin-press-var}.
\begin{equation}
\begin{aligned}
\frac{Q^\text{Linear Pressure}_\text{Deformable}}{Q_\text{Rigid}}&=1+d_1\chi+d_2{\chi}^2+d_3{\chi}^3\\
d_1&=\frac{4}{5}=0.8\\
d_2&=\frac{16}{75}\approx 0.2133\\
d_3&=\frac{64}{3375}\approx 0.0190
\end{aligned}
\label{Eq-vfr-lin-press-var}
\end{equation}
In section~\ref{Modeling_results_Linear pressure assumption},
a criterion is sought to identify the threshold value separating the
two regimes of coupled and decoupled fluid-solid mechanics through examining the Eqs.~\ref{Eq-sol-gen-Non-dim-straight-Q-flex-over-Q-rig-analytical} and \ref{Eq-vfr-lin-press-var}.

\section[Experiments]{Experiments}
\label{Experiments}

\subsection[Fabrication]{Fabrication}
\label{Fabrication} 
\begin{figure}[t!]
\centering
	\subfloat{}{%
	\includegraphics[width=0.7\linewidth]{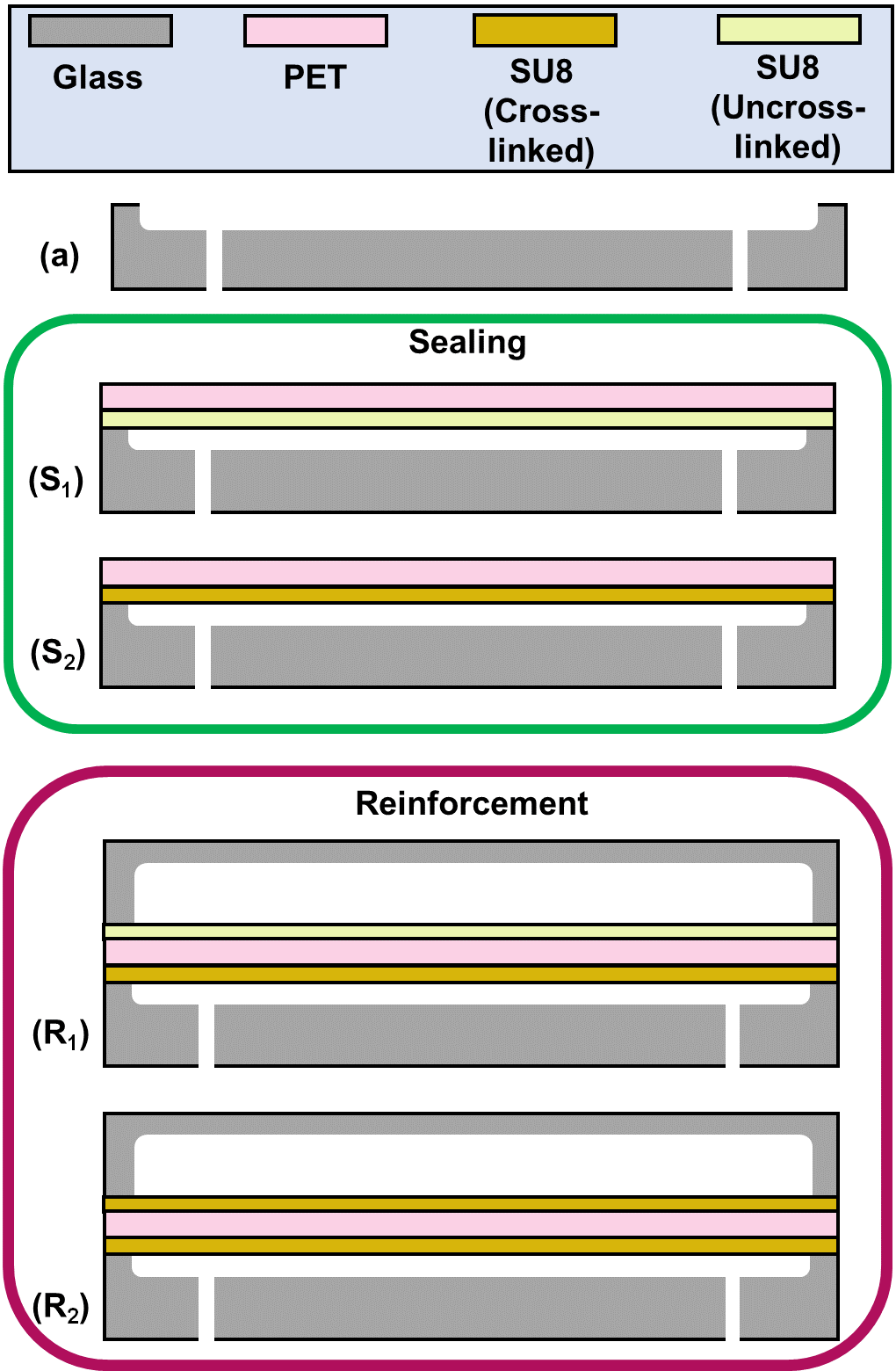}
	}
	\caption{
	Schematic representation of the fabrication process flow:
	(a) the microchannel sidewalls are created using the double-step glass wet etching.
	(S\textsubscript{1}) A PET film coated with a thin layer of SU8 seals the microchannel sidewalls.
	(S\textsubscript{2}) The chip is undergone with UV illumination and post-exposure baking to cure the adhesive layer.
	(R\textsubscript{1}) A glass substrate with the same pattern as that of the main channel is aligned 
	and brought in contact with the SU8 layer coated on the bare side of the PET film.
	(R\textsubscript{2}) The sample is undergone with UV illumination and post-exposure baking to cure the adhesive layer at the bonding interface of the reinforcement glass and the PET film.
	}
	\label{Fig-Fab}
\end{figure}

\begin{figure}[t!]
\centering
	\subfloat{}{%
	\includegraphics[width=1\linewidth]{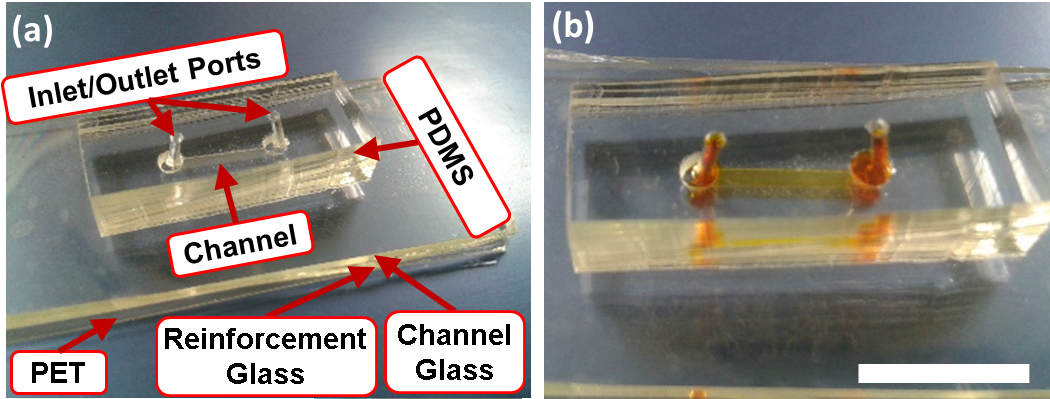}
	}
	\caption{
	(a) A photograph of the fabricated device together with (b) a photograph showing the channel filled with a dye demonstrating a leakage-fee bonding. The scale bar is $1~cm$ in (b).
	}
	\label{Fig-Device}
\end{figure}

The schematic representation of the fabrication process flow is shown in Fig.~\ref{Fig-Fab}.
The procedure is similar to what we described in our previous work \cite{mehboudi_two-step_2018} with the difference that the flexible membrane is allowed to deform even after the reinforcement step. 
Briefly, the patterns of the microchannel sidewalls are created through the glass wet etching procedure.
The via-holes are drilled in the glass slide using a 2-mm drill bit; see Fig.~\ref{Fig-Fab} (a).
The $100~\mu m$-thick PET (Polyethylene terephthalate) film coated with a thin SU8 film ($\sim 2~\mu m$) then seals the channel sidewalls on the hotplate at $69\pm 1 ^\circ C$;
see Fig.~\ref{Fig-Fab} (S\textsubscript{1}).
After 30 minutes, the temperature decreases to the room temperature over 20 minutes. 
The SU8 adhesive layer is cured using the UV exposure of $\sim 1.3~J/cm^2$ and baking on the hotplate, initially at $65^\circ C$ ramping up to $95^\circ C$ over 5 minutes, remaining at this temperature for 30 minutes and cooling down to the room temperature over 30 minutes; see Fig.~\ref{Fig-Fab} (S\textsubscript{2}).
A separate glass slide is patterned and etched to create a cavity for accommodating the deflected membrane while providing the bonding reinforcement; see Fig.~\ref{Fig-Fab} (R\textsubscript{1}). 
The etched cavity is sufficiently deep ($\geq 45~\mu m$) to enable the membrane to displace freely.
The reinforcement glass slide is bonded with the PET/glass microchannel stack using the partially cured SU8 as an adhesive layer.
Prior to bonding, the SU8 layer of about $14~\mu m$ in thickness is spun onto the bare side of the PET film and baked on the hotplate at $65^\circ C$ for 4 minutes.
During bonding, the stack is baked at $69^\circ C$ for 20 minutes, followed by the UV exposure of $2~J/cm^2$ and postbaking at $95^\circ C$ for 30 minutes; see Fig.~\ref{Fig-Fab} (R\textsubscript{2}).
A PDMS interconnect with holes are bonded to the glass slides for injection and collection of the fluid.
The photographs of a fabricated device are shown in Fig.~\ref{Fig-Device}.
A dye was injected through the channel to ensure a leak-free bonding.
The photograph shown in Fig.~\ref{Fig-Device} (b) demonstrates a leakage-free bonding between the PET film and the glass as well as between the PDMS and the glass.

\subsection[Volumetric flow rate measurement]{Volumetric flow rate measurement}
\label{Volumetric flow rate measurement} 
In order to create a pressure driven flow under a specific pressure difference, a reservoir is used with two ports.
The compressed air, the pressure of which is adjusted using a pressure regulator (PneumaticPlus, PPR2-N02BG-4 Miniature Air Pressure Regulator), is injected through one of the ports causing the liquid to be pushed out from the other port, which is connected to the microchannel's inlet.
Two approaches are used to measure the volumetric flow rate depending on its magnitude.

\subsubsection[Low flow rate regime]{Low flow rate regime (below about 40 micro-litre/min)}
\label{Low flow rate regime}
The discharged liquid is guided through a capillary tube.
The meniscus displacement, $\Delta x_m$, and the time interval, $\Delta t_m$, are measured over 2--7 centimeters of the capillary tube length depending on the meniscus speed.
The meniscus speed is calculated from $u_m=\Delta x_m/\Delta t_m$.
The volumetric flow rate equals $\frac{\pi}{4} D^2_c u_m$, where $D_c$ refers to the capillary inner diameter.
Two different capillary tubes with $D_c=$ 530 and 793.75 microns were used for the volumetric flow rates below $\sim 2~\mu l/min$ and above that, respectively.
The hydrodynamic resistance of the capillary tubes ($<10^{11} Pa.s/m^3$) is at least three orders of magnitude smaller than that of the studied devices to avoid the spurious effects on the measured volumetric flow rate.

\subsubsection[High flow rate regime]{High flow rate regime (above about 40 micro-litre/min)}
\label{High flow rate regime} 
The liquid discharging from the microchannel outlet is collected in a vial, the mass of which is monitored over time using an analytical balance (Mettler Toledo-XS105, 0.01 milligram resolution).
The slope of the accumulated mass versus time is the mass flow rate, $\dot{m}$.
The volumetric flow rate is obtained from $Q=\dot{m}/\rho$, where $\rho$ denotes the fluid density.

\section[Results and discussion]{Results and discussion}
\label{Results and discussion} 

\subsection[Parametric study of the flexibility parameter]{Parametric study of the flexibility parameter}
\label{Modeling_results}

\begin{figure}[t!]
	\centering
	\subfloat{
	\includegraphics[width=1\linewidth]{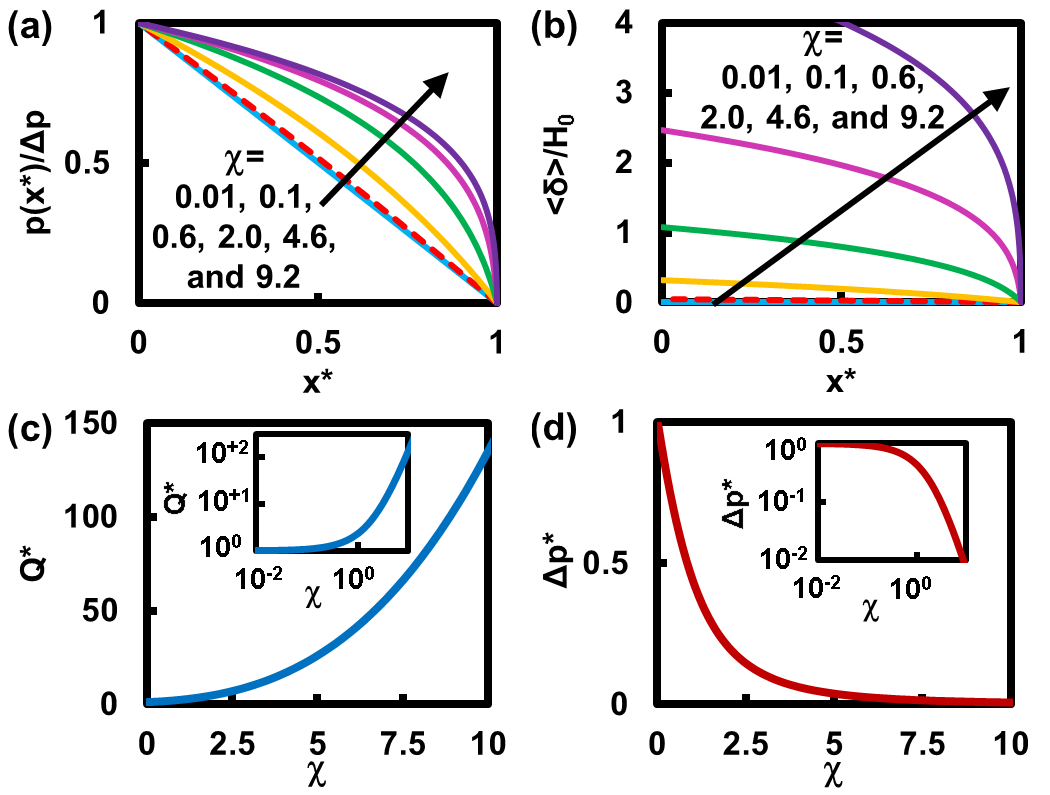}
	}
	\caption{(a) Plots of the dimensionless pressure, $p^\ast=p(x^\ast)/\Delta p$, as a function of the dimensionless distance from the inlet $x^\ast$ for various flexibility parameter values \cite{christov_flow_2018}.
	(b) The variation of the dimensionless membrane's mean deflection, 
	$<\delta>/H_0$, 
	along the microchannels for various flexibility parameter values.
	(c) A plot of the dimensionless flow rate $Q^\ast=\big [{Q_\text{Deformable}}/{Q_\text{Rigid}}\big ]_{\Delta p_\text{Deformable}=\Delta p_\text{Rigid}}$, as a function of the flexibility parameter. 
	(d) A plot of the dimensionless pressure difference $\Delta p^\ast=\big [\Delta p_\text{Deformable}/\Delta p_\text{Rigid}\big ]_{Q_\text{Deformable}=Q_\text{Rigid}}$, as a function of the flexibility parameter.
	The insets show the logarithmic plots in (c) and (d).
	}
	\label{Fig-Straight_pressure_vfr-00}
\end{figure}

The dimensionless pressure variation along the deformable microchannel is plotted for various flexibility parameters as shown in Fig.~\ref{Fig-Straight_pressure_vfr-00} (a).
It is perceived that the pressure variation deviates more from the linear one as the flexibility parameter increases.
In case of the large flexibility parameters, the slope of the curve increases in magnitude towards the exit of the channel (increasing $x^\ast$), which can be attributed to the increasing local hydrodynamic resistance. 
The larger pressure levels close to the inlet cause the larger membrane deflections, decreasing the hydrodynamic resistance through increasing the cross-sectional area; see Fig.~\ref{Fig-Straight_pressure_vfr-00} (b).
From Fig.~\ref{Fig-Straight_pressure_vfr-00} (b), at any $x^\ast$, the bulging becomes larger as the flexibility parameter increases, inducing the more non-linear pressure variation within the microchannel, as shown in Fig.~\ref{Fig-Straight_pressure_vfr-00} (a).

Plots of the dimensionless flow rate ($Q^\ast$) and dimensionless pressure drop ($\Delta p^\ast$) are shown in Fig.~\ref{Fig-Straight_pressure_vfr-00} (c) and (d) as a function of $\chi$ and can be directly obtained from Eq.~\ref{Eq-sol-gen-Non-dim-straight-Q-flex-over-Q-rig-analytical}.
While $Q^\ast$ is a cubic polynomial function of $\chi$, $\Delta p^\ast$ is a rational function, {\it i.e.} $\Delta p^\ast = 1/Q^\ast$.
The {\it dynamic rescaling} of the vertical axis takes place when plotting 
$Q^\ast$ versus $\chi$ because of the term of $\Delta p$ included in the definition of $Q^\ast$, allowing us to obtain the master curve shown in Fig.~\ref{Fig-Straight_pressure_vfr-00} (c). 
The vertical axis represents an extent of deviation between the characteristic behavior of a deformable microchannel and that of its theoretically rigid counterpart.

Figure~\ref{Fig-Straight_pressure_vfr-00} (d)
shows what fraction of the pressure difference applied across its rigid counterpart is needed to be applied across a deformable microchannel to produce the same volumetric flow rate.
It is perceived that for large values of $\chi$, small fractions of the pressure difference 
are needed to be applied across the deformable microchannels to deliver the same volumetric flow rates as those of their rigid counterparts.

\subsection[Experimental results]{Experimental results}
\label{Flexible_Straight_Results_Experimental_modeling}
Four different shallow microchannels with ultra-low height-to-width ratios have been fabricated.
The microchannels geometries are summarized in Table~\ref{tbl_channel_geom}.
\begin{table*}[t]
\small
  \caption{The geometry of four deformable microchannels studied in this paper.}
  \label{tbl_channel_geom}
	\begin{tabular*}{1.0\textwidth}{@{\extracolsep{\fill}}lllll}
    \hline
    Microchannel  & Length ($mm$) & Width ($mm$) & Height ($\mu m$) & Height-to-Width Ratio\\
    \hline
		{\bf A} & $9.5\pm 0.5$ & $1.25\pm 0.1$ & $2.3\pm 0.1$ & $1.84\times 10^{-3}$\\
    {\bf B} & $9.5\pm 0.5$ & $2.0\pm 0.1$ & $2.3\pm 0.1$ & $1.15\times 10^{-3}$\\
		{\bf C} & $9.5\pm 0.5$ & $1.25\pm 0.1$ & $7.5\pm 0.4$ & $6.00\times 10^{-3}$\\
		{\bf D} & $9.5\pm 0.5$ & $2.2\pm 0.1$ & $8.2\pm 0.4$ & $3.73\times 10^{-3}$\\
    \hline
  \end{tabular*}
\end{table*}
In our mathematical modeling, the modulus of elasticity and Poisson's ratio of the PET are 2.5 GPa and $0.4$, respectively.
The modulus of elasticity of the SU8 is considered to be 2 GPa (MicroChem Corp.).
Although the Poisson's ratio of SU8 is not needed in this work, for the sake of completeness, this property is $\sim 0.2$--0.34 \cite{luo_simple_2003}.
Since the thicknesses of the SU8 layers and the PET film are t\textsubscript{s1}$=2~\mu m$, t\textsubscript{s2}$=14~\mu m$, and t\textsubscript{p}$=100~\mu m$ (Fig.~\ref{Fig-Schem-transformed-section}), the equivalent thickness of the SU8-coated PET membrane equals $\tilde{t}\approx 113.3~\mu m$.
The thermophysical properties of the DI water are considered to be $\rho=1000~kg/m^3$ and $\mu=8.9\times 10^{-4}~Pa.s$.

\begin{figure*}[ht]
\centering
	\subfloat{%
	\includegraphics[width=1\textwidth]{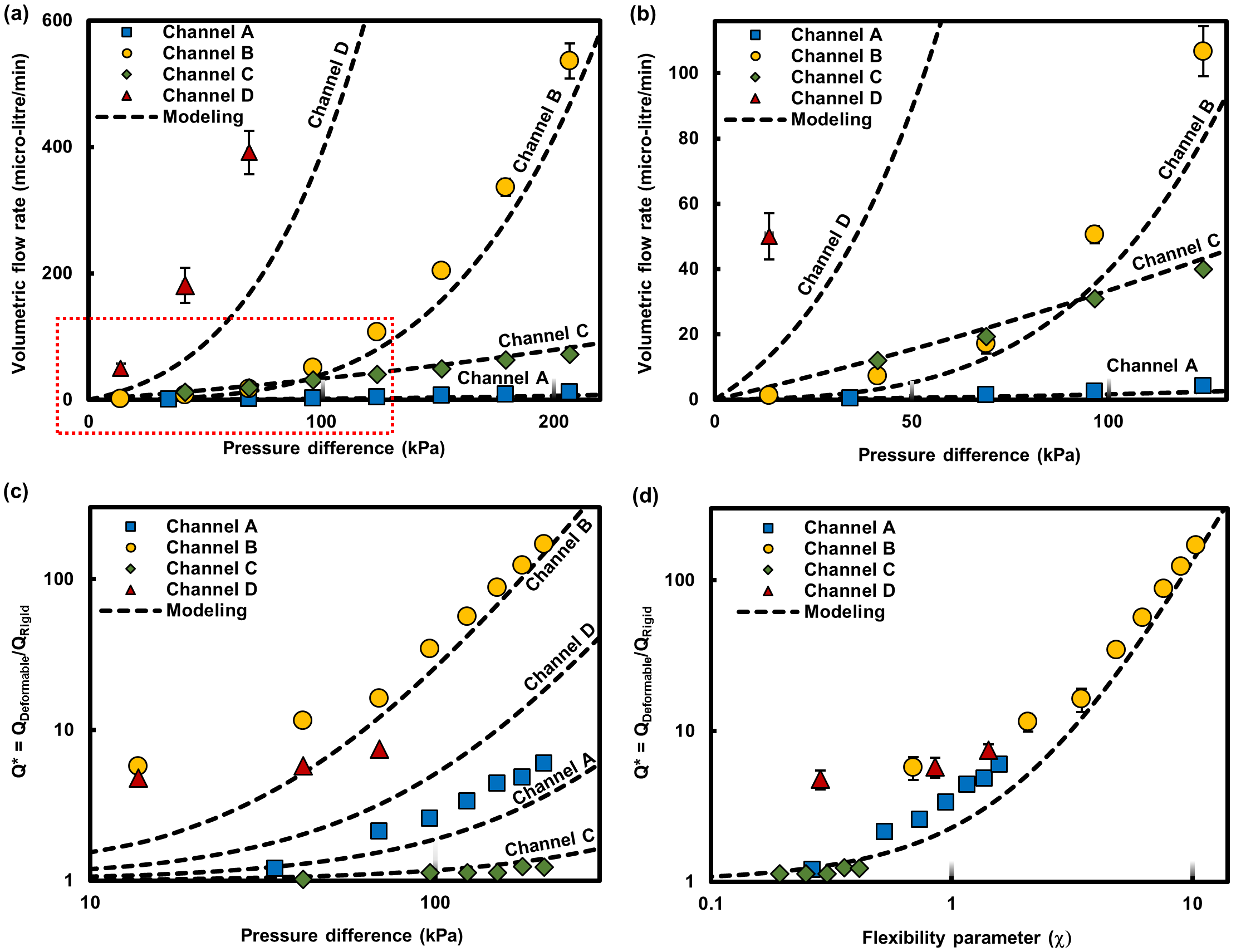}
	}
	\caption{(a) The volumetric flow rate through Microchannels A, B, C, and D in comparison with the analytical solutions of the one-dimensional model~\cite{christov_flow_2018}~together with (b) the magnification of the dotted box.
	(c) The ratio of the volumetric flow rate through the deformable microchannels to that through their theoretically rigid counterparts under the same pressure differences. 
	(d) The master curve obtained from Eq.~\ref{Eq-sol-gen-Non-dim-straight-Q-flex-over-Q-rig-analytical}~together with the experimental results.
	A logarithmic scale is used in (c) and (d).
	The error bars, obtained from three data-sets, are not shown when they are smaller than the size of markers.
	}
	\label{Fig-Res-vfr-ch-ABCD}
\end{figure*}

The volumetric flow rates through Microchannels A, B, C, and D are shown in Fig.~\ref{Fig-Res-vfr-ch-ABCD} (a) and (b) for various pressure differences across the microchannels.
The volumetric flow rates of Microchannels A--D normalized by those of the rigid counterparts, {\it i.e.} $Q^\ast={Q_\text{Deformable}}/{Q_\text{Rigid}}$, are also shown in Fig.~\ref{Fig-Res-vfr-ch-ABCD}~(c) under various pressure drops.
A good agreement is observed between the one-dimensional model developed by Christov {\it et. al.} \cite{christov_flow_2018} and the experimental results.
The larger discrepancies between the modeling and experimental results of Microchannels B and D can be attributed to the rather large width-to-length ratio, $W/L\approx 0.22$, which does not perfectly satisfy the requirement of the wide-beam framework and lubrication theory, {\it i.e.} $W/L\ll 1$.
If the microchannel is not sufficiently long, the membrane's infinitesimal slices normal to the flow-wise direction cannot be accurately approximated by the wide beams.
Under these situations, at any $x^\ast$, the membrane deflection is not solely dependent on the corresponding local pressure $p(x^\ast)$ within the channel.
Instead, the local membrane deflection is influenced by the upstream pressure as well.
Therefore, the membrane displacement becomes larger than predicted by the model based on the wide-beam framework and lubrication theory, causing the model to underestimate the volumetric flow rate.

In addition to the larger value of $W/L$ for Microchannel D ($\approx 0.232$) compared with that of Microchannel B ($\approx 0.211$), the larger discrepancies from modeling observed in experimental mass flow rates through Microchannel D may be partially due to the inaccurate location of the inlet/outlet via-holes drilled manually in glass slides.
In our fabricated devices, the center of the holes has an offset of 0.1--0.5 $mm$ from the expected locations, 
causing uncertainties to rise not only in the overall length of microchannels, but also in the channels' width in proximity of the inlet/outlet ports; see Fig.~\ref{Fig-Device}~(b). 
In case of narrow microchannels, {\it e.g.} a few hundreds of microns in width, the interface between the main part of channel and the inlet/outlet via-holes of $2~mm$ in diameter can be reasonably assumed to be a straight line. 
Under these situations, the channel width remains intact after creating the via-holes.
Whereas, for wide microchannels like Microchannels B and D, the round shape of the inlet/outlet via-holes can significantly change the microchannel width close to the inlet/outlet ports.
For such situations, the channel's width is not constant in the proximity of the inlet/outlet sections. 
An offset from the expected locations of the via-holes causes even more deviations from the expected straight microchannel. 
Therefore, the fluid flows through Microchannels B and D can be quite different from each other close to the inlet/outlet sections due to the existing lateral and axial offsets in locations of the created via-holes.

Figure~\ref{Fig-Res-vfr-ch-ABCD}~(d) shows that, to a reasonable agreement, the experimental data points of flow rate versus pressure drop for all four channels fall onto the master curve presented in Eq.~\ref{Eq-sol-gen-Non-dim-straight-Q-flex-over-Q-rig-analytical}~over a large range of the flexibility parameter.
This master curve demonstrates that  
difference between
the volumetric flow rate of a deformable channel and that of its rigid counterpart under the same pressure drop is solely dictated by the flexibility parameter introduced in this paper.
Because of having a relatively large width and small height, Microchannel B exhibits the largest flexibility parameter and consequently the most significant 
deviation from
its rigid counterpart (largest $Q^\ast$), whereas the relatively small width and large height of Microchannel C cause it to show the smallest flexibility parameter and consequently the closest resemblance to its rigid counterpart (smallest $Q^\ast$). 
In the following, we are elaborating how the flexibility parameter's scale dictates the characteristic behavior of the fluid flow through deformable microchannels.

According to Fig.~\ref{Fig-Res-vfr-ch-ABCD} (a), under the sufficiently large pressure differences, Microchannels B and D deliver much larger volumetric flow rates than Microchannels A and C because the larger membrane deflection and averaged cross-section area result from the larger width of Microchannel B and D.
In particular, under the maximum investigated pressure, {\it i.e.} $\sim 207$ kPa, the volumetric flow rate delivered by Microchannel B is about 70 times larger than that of Microchannel A.
It is worth mentioning that such distinct characteristic behaviors emerge due to 
the rather small difference of the microchannels widths, ${(W_B-W_A)}/{W_A}\approx 0.6$.

According to Fig.~\ref{Fig-Res-vfr-ch-ABCD} (c),
the deformable microchannels behave similar to their corresponding rigid microchannels $(Q^\ast\approx 1)$ under the sufficiently low pressure differences.
By increasing the pressure difference, however, $Q^\ast$ increases and the difference between the deformable and rigid microchannels becomes more significant.
The graphs also show that  $Q^\ast$ increases as the microchannel width increases and/or its height decreases.
Compared to its corresponding rigid microchannel, Microchannel C shows only $40\%$ increase in the volumetric flow rate under 207 kPa, while Microchannel B delivers the flow rate of about 148 times larger than its corresponding rigid microchannel.
Despite the fact that Microchannel C's height is more than three times as large as that of Microchannel B, the volumetric flow rate delivered by Microchannel B under 207 kPa is about six times as large as that delivered by Microchannel C.
A rigid microchannel
possessing triple a specific height would result in a 26-time larger volumetric flow rate 
due to the term of $H^3_0$ in $Q_\text{Rigid}=\Delta pWH^3_0/12\mu L$.
It is worth mentioning that under the sufficiently low pressure differences, where the fluid flow characteristics are more similar to those of the rigid microchannel, Microchannel C delivers the larger volumetric flow rate in comparison with Microchannel B; see Fig.~\ref{Fig-Res-vfr-ch-ABCD} (b).
By increasing the pressure difference, the volumetric flow rate of Microchannel B surpasses that of Microchannel C at $\Delta p\approx 92$ kPa.
In order to obtain a more profound insight into 
the underlying physics of the observed distinct flow characteristics, 
the Eq.~\ref{Eq-sol-gen-Non-dim-straight-Q-flex-over-Q-rig-analytical} is more deeply examined in the following.

The order of magnitude of different terms existing in Eq.~\ref{Eq-sol-gen-Non-dim-straight-Q-flex-over-Q-rig-analytical} is presented in Table~\ref{tbl_terms_order} for the flexibility parameter, $\chi$, with various orders of magnitude.
\begin{table*}[t]
\small
  \caption{The order of magnitude of different terms existing in Eq.~\ref{Eq-sol-gen-Non-dim-straight-Q-flex-over-Q-rig-analytical} for the flexibility parameter, $\chi$, with various orders of magnitude.
	For each flexibility scale, the most important terms with the order of magnitude equal to or greater than $10^{-1}\times O(Q^\ast)$ are 
	highlighted in bold font.
	Note: $O(c_1)=1$, $O(c_2)=10^{-1}$, and $O(c_3)=10^{-1}$.}
  \label{tbl_terms_order}
	\begin{tabular*}{1.0\textwidth}{@{\extracolsep{\fill}}llllll}
    \hline
      $O(\chi)$ & $10^{-2}$ & $10^{-1}$ & $10^{0}$ & $10^{+1}$ & $10^{+2}$\\
    \hline
		\hline
		$O(1)$ & $\bm{10^{0}}$ & $\bm{10^{0}}$ & $\bm{10^{0}}$ & $10^{0}$ & $10^{0}$\\
		$O(c_1\chi)$ & $10^{-2}$ & $\bm{10^{-1}}$ & $\bm{10^{0}}$ & $\bm{10^{+1}}$ & $10^{+2}$\\
		$O(c_2\chi^2)$ & $10^{-5}$ & $10^{-3}$ & $\bm{10^{-1}}$ & $\bm{10^{+1}}$ & $10^{+3}$\\
		$O(c_3\chi^3)$ & $10^{-7}$ & $10^{-4}$ & $\bm{10^{-1}}$ & $\bm{10^{+2}}$ & $\bm{10^{+5}}$\\
    \hline
  \end{tabular*}
\end{table*}
This table shows the contribution of each term in the volumetric flow rate through the deformable microchannels. 
The terms with $10\%$ contribution or more are highlighted in bold font for each order of magnitude of $\chi$.
It is perceived that the higher orders of the flexibility parameter become more important as the flexibility parameter increases.
The magnitude of the truncation error can also be extracted from this table
for the expressions consisting of various terms,
that can be used to approximate the Eq.~\ref{Eq-sol-gen-Non-dim-straight-Q-flex-over-Q-rig-analytical}. 
The important finding is that a variety of distinct characteristic behaviors can emerge under different flexibility parameter scales.
For example, the dimensionless volumetric flow rate can be approximated as $Q^\ast\approx 1$ and $Q^\ast\approx c_3\chi^3$ for the flexibility parameter scales of $O(\chi)\leq 10^{-2}$ and $O(\chi)\geq 10^{+2}$, respectively, while the truncation error is only about $1\%$ of the exact value.
For the same truncation error, Eq.~\ref{Eq-sol-gen-Non-dim-straight-Q-flex-over-Q-rig-analytical} can be approximated as 
$Q^\ast\approx 1+c_1\chi$  and 
$Q^\ast\approx c_1\chi+c_2\chi^2+c_3\chi^3$ 
 for the flexibility parameter scales of $O(\chi)= 10^{-1}$ and $O(\chi)= 10^{+1}$, respectively, while all the terms are needed for the flexibility scale of $O(\chi)= 10^{0}$ and truncation error of $1\%$.

From the definition of the flexibility parameter, $\chi={\Delta pW^4}/{384DH_0}$, 
the fluid-solid behavioral characteristics rely on the following three general categories:
1) microchannel dimensions: $W$ and $H_0$,
2) membrane thickness and structural properties, which are lumped into the parameter of $D$, and
3) pressure difference across the microchannel, $\Delta p$.
Regardless of the microchannel geometry and the membrane properties, 
if the pressure difference is {\bf sufficiently} low,
the flexibility parameter becomes much smaller than unity.
Under these situations,
the first term in the right side of Eq.~\ref{Eq-sol-gen-Non-dim-straight-Q-flex-over-Q-rig-analytical} is dominant:
${Q^\ast}\approx 1$.
As long as the condition of $\chi\ll 1$ is satisfied, any change in the dimensions of the microchannel, pressure difference, or membrane properties causes no significant influence on $Q^\ast$, and the deformable microchannel behaves similar to its corresponding rigid microchannel.
On the other hand, for the {\bf sufficiently} large pressure difference across the flexible microchannels, the last term becomes dominant:
$Q^\ast\approx c_3{\chi}^3$.
Under these situations, a slight change in the microchannel dimensions, pressure difference, or membrane properties is significantly magnified because of the term of ${\chi}^3$.
As a result, two deformable microchannels
slightly different in their dimensions and/or the membrane properties might exhibit the similar or the noticeably different 
characteristic behaviors dependent upon the pressure difference applied across the microchannels.

A sufficiently large change in the pressure difference across a specific fabricated microchannel 
can activate/inactivate the different terms in the right side of Eq.~\ref{Eq-sol-gen-Non-dim-straight-Q-flex-over-Q-rig-analytical} through changing the order of magnitude of the flexibility parameter; see Table~\ref{tbl_terms_order}.
As a result, a deformable microchannel might exhibit a variety of distinct characteristic behaviors depending on the pressure difference applied across the microchannel.

\begin{figure*}[ht]
\centering
	\subfloat{
	\includegraphics[width=1\linewidth]{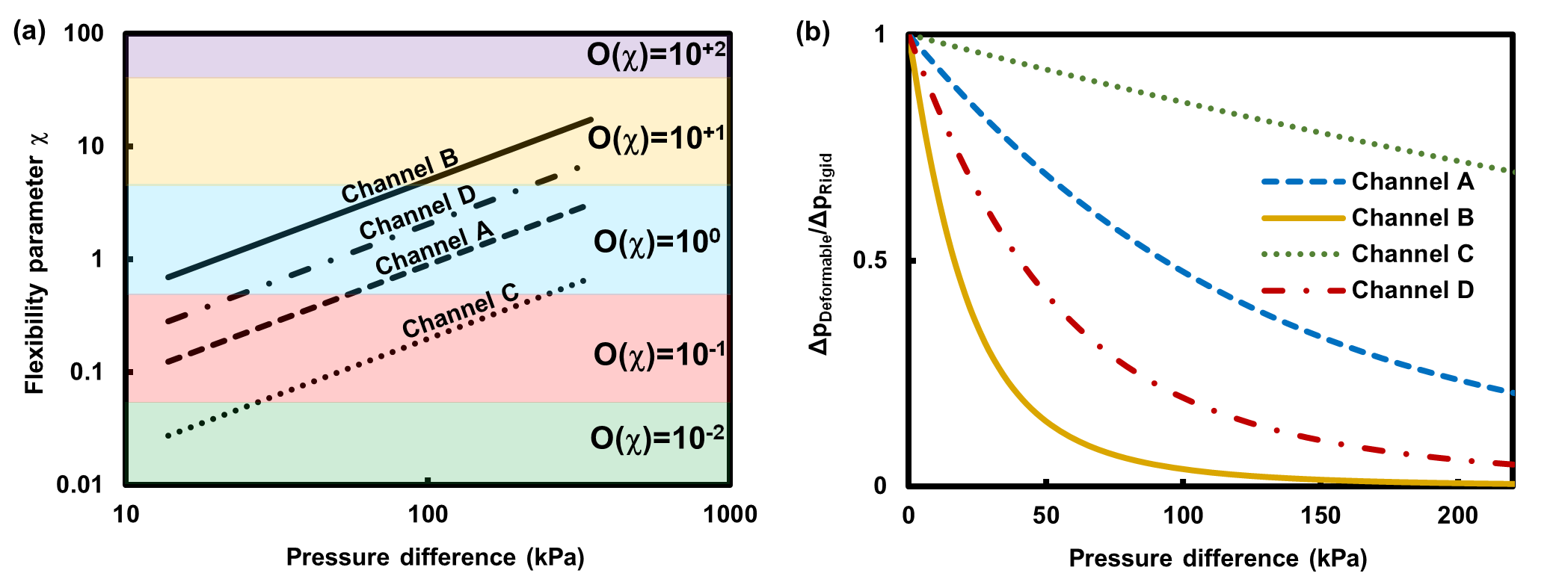}
	}
	\caption{
	(a) The flexibility parameter, $\chi$, for Microchannels A, B, C and D under various pressure differences.
	The differences between the channels geometries and/or the membranes properties result in the different flexibility parameters under the same pressure difference across the channels.
		A sufficiently large change in the pressure difference across a specific channel can alter the flexibility parameter scale from one characteristic regime to another.
	(b) The fraction of the pressure difference across the rigid microchannel sufficiently large to cause the same volumetric flow rate when applied across the deformable microchannel with the same dimensions.
	}
	\label{Fig-Res-chi-modified}
\end{figure*}

The flexibility parameter is presented in Fig.~\ref{Fig-Res-chi-modified} (a) as a function of the pressure difference across the investigated microchannels.
Even though the flexibility parameter of the microchannels varies with the applied pressure difference, 
the ratio of the flexibility parameters related to the different microchannels of $i$ and $j$ remains constant under various applied pressure differences: 
\begin{equation}
\bigg[\chi_i/\chi_j\bigg]_{\Delta p_i=\Delta p_j}=(W_i/W_j)^4\times (H_{0,j}/H_{0,i})
\label{Eq-chi-mod-AB}
\end{equation}
where $W_i$ and $W_j$ refer to the widths of the microchannels $i$ and $j$, respectively.
Similarly, $H_{0,i}$ and $H_{0,j}$ show the original heights of the microchannels $i$ and $j$, respectively.
Because of the term of $W^4$, any difference in width of the microchannels causes a noticeable difference in the flexibility parameters and potentially the fluid flow characteristics.
This is the reason that a $60\%$ larger width results in Microchannel B's flexibility to be 
$1.6^4\approx 6.6$
times as large as that of Microchannel A.
This ratio is sufficiently large to cause the two microchannels to operate within the different flexibility regimes.
For example, under the pressure difference of 207 kPa, where $\chi_A\approx 1.59$ and $\chi_B\approx 10.39$, the order of magnitude of $\chi$ is $10^{0}$ and $10^{+1}$ for Microchannels A and B, respectively.
As a result, the term of $c_3\chi^3$ plays the most important role in the characteristic behavior of Microchannel B (Table~\ref{tbl_terms_order}), while the lower orders of $\chi$ dictate the behavior of Microchannel A, which causes the volumetric flow rate through Microchannel B to be about 70 times larger than that of Microchannel A.

According to Table~\ref{tbl_terms_order}, for the sufficiently small flexibility parameter, {\it e.g.} $O(\chi)\leq 10^{-1}$, the term of $1$ is dominant in Eq.~\ref{Eq-sol-gen-Non-dim-straight-Q-flex-over-Q-rig-analytical}.
The volumetric flow rate is, therefore, mainly proportional to $WH^3_0$, which resembles the rigid microchannel characteristics.
By contrast, for the large-flexibility regime, {\it e.g.} $O(\chi)\geq 10^{+1}$, the term of $c_3\chi^3$ plays the most important role.
In these situations, the volumetric flow rate is proportional to $W^{13}$.
One can show that
\begin{equation}
\begin{aligned}
\chi\gg 1~:~{Q_\text{Deformable}}\approx \frac{1}{7,970,586,624}\times\frac{W^{13}{\Delta p}^4}{\mu L{D}^3}.
\end{aligned}
\label{Eq-large-flex-Q-1}
\end{equation}
It is perceived that the height-dependency vanishes, while the microchannel width plays a significant role in the fluid-solid characteristic behavior.
Despite the fact that the height of Microchannel C is about 3.3 times as large as that of Microchannel B, 
such distinct characteristics cause the volumetric flow rate through Microchannel B to be about six times as large as that through Microchannel C under 207 kPa (Fig.~\ref{Fig-Res-vfr-ch-ABCD} (a)), whereas the volumetric flow rate through Microchannel C is about 8 times larger than that through Microchannel B under the relatively low pressure difference of 41 kPa (Fig.~\ref{Fig-Res-vfr-ch-ABCD} (b)).
At the lowest extreme of the pressure range, where both microchannels behave similar to their corresponding rigid microchannels, we have $Q_C/Q_B=W_C/W_B\times (H_{0,C}/H_{0,B})^3\approx 21.7$.

The fraction of the pressure difference across the rigid microchannels enough to produce the same volumetric flow rate when applied across the corresponding deformable microchannels is shown in Fig.~\ref{Fig-Res-chi-modified} (b).
Microchannel B exhibits the smallest fraction values because of having the largest flexibility parameter.  
Under the pressure difference of 207 kPa, $\Delta p^\ast$
is about $6.7\times 10^{-3}$ for Microchannel B, which means a rigid microchannel with the same dimensions as those of the undeformed Microchannel B would need about 149 times as large pressure as 207 kPa, {\it i.e.} $\approx 30.8$ MPa, to deliver the same volumetric flow rate 
($\sim 468$ micro-litre/min).

\begin{figure*}[ht]
\centering
	\subfloat{
	\includegraphics[width=1\linewidth]{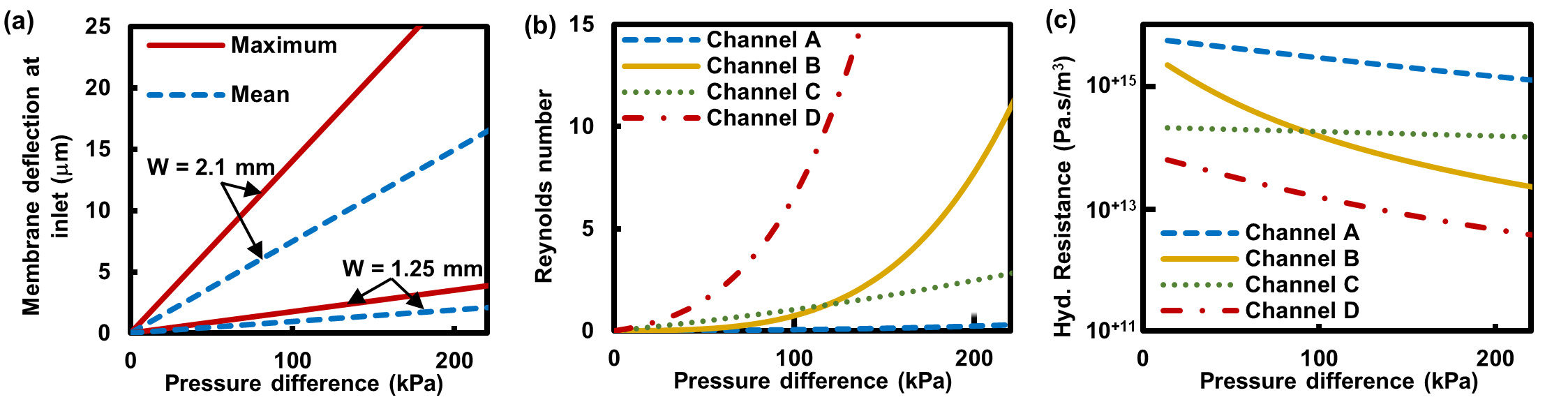}
	}
	\caption{
	(a) The mean and maximum membrane deflections at the inlet section of Microchannels A and C (W=1.25 mm) as well as those for Microchannels B and D (W=2.1 mm) {obtained using the theoretical model} \cite{christov_flow_2018}.
	(b) The Reynolds number and (c) the hydrodynamic resistance for the studied microchannels under various pressure differences.
	}
	\label{Fig-Res-deflection-Re-Resist}
\end{figure*}

The mean and maximum deflections of the membrane obtained using the theoretical model \cite{christov_flow_2018} are shown in Fig.~\ref{Fig-Res-deflection-Re-Resist} (a) for the inlet section.
The wider microchannels have about 0.6 times larger width, which is sufficiently large to cause the noticeably larger deflections, since the membrane deflection is proportional to $W^4$ \cite{christov_flow_2018}.
The maximum membrane deflection is reasonably smaller than the membrane thickness ($100~\mu m$), which suggests that the {\it thin membrane bending} assumption is valid \cite{christov_flow_2018}.
The Reynolds number is also calculated for the microchannels:
$\text{Re}={\rho \bar{u}D_{h}}/{\mu}={2\rho Q}/{\mu W}$,
in which $\bar{u}$ and $D_{h}$ denote the average velocity and the hydraulic diameter of the microchannel, respectively, where $D_{h}\approx2H$.
The results are shown in Fig.~\ref{Fig-Res-deflection-Re-Resist} (b).
The assumption of the laminar flow regime is clearly valid in this study.
In addition, the hydrodynamic resistance, $R_h=\Delta p/Q$, of the microchannels is presented in Fig.~\ref{Fig-Res-deflection-Re-Resist} (c).
Microchannel C shows the least variations of the hydrodynamic resistance due to the small membrane deflections compared to its original height, whereas Microchannel B exhibits a-few-orders-of-magnitude change in the hydrodynamic resistance over the investigated pressure difference range.

\subsection[Linear pressure variation assumption]{Linear pressure variation assumption}
\label{Modeling_results_Linear pressure assumption}

\begin{figure}[ht]
\centering
	\subfloat{
	\includegraphics[width=1\linewidth]{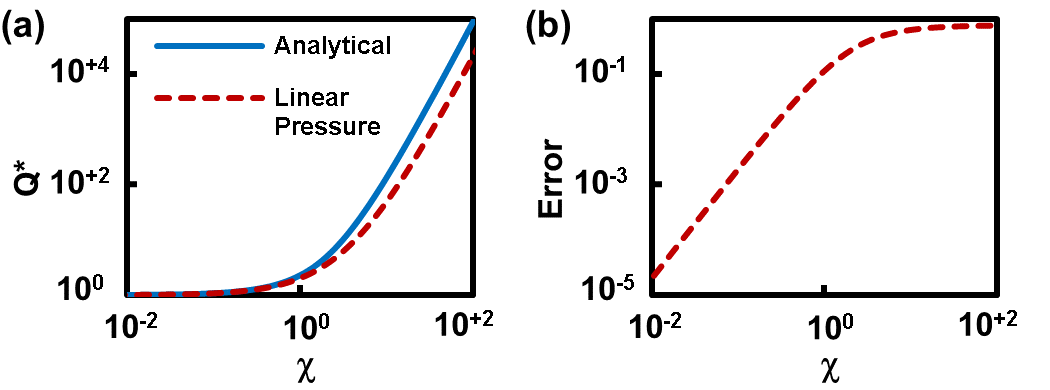}
	}
	\caption{(a) The dimensionless volumetric flow obtained through the linear pressure variation assumption, Eq.~\ref{Eq-vfr-lin-press-var}, compared with that obtained through the analytical solution \cite{christov_flow_2018}, Eq.~\ref{Eq-sol-gen-Non-dim-straight-Q-flex-over-Q-rig-analytical}, as a function of the flexibility parameter.
	(b) The error associated with the linear pressure variation assumption:
	$\big(Q^\text{Analytical}_\text{Deformable}-Q^\text{Linear Pressure}_\text{Deformable}\big)/Q^\text{Analytical}_\text{Deformable}$.
	}
	\label{Fig-Res-lin-press}
\end{figure}

The dimensionless volumetric flow rate related to the one-dimensional decoupled fluid-solid mechanics model based on the linear pressure variation assumption, Eq.~\ref{Eq-vfr-lin-press-var}, is shown in Fig.~\ref{Fig-Res-lin-press} (a) in comparison with the analytical one-dimensional coupled fluid-solid mechanics model \cite{christov_flow_2018}, Eq.~\ref{Eq-sol-gen-Non-dim-straight-Q-flex-over-Q-rig-analytical}. 
The volumetric flow rate is underestimated by assuming the linear pressure variation through the microchannel.
The associated error is shown in Fig.~\ref{Fig-Res-lin-press} (b).
It is observed that the linear pressure assumption is acceptable for the small-flexibility regime.
In particular, the error is less than $1\%$ if $\chi<0.241$.
From Fig.~\ref{Fig-Straight_pressure_vfr-00} (a), one can perceive that the pressure behavior obtained through the analytical solution is close to the linear variation for such a small flexibility parameter, which explains the acceptable agreement between the coupled and decoupled models.
The flexibility parameter for many of the deformable microchannel applications is smaller than $0.241$.
The maximum flexibility parameter investigated in the comparative study of several benchmarks in \cite{shidhore_static_2018} equals $\chi\approx 0.209$.
As previously explained, our work allowed us to study the low-Reynolds-number fluid flow in the deformable microchannel with the new behavioral characteristic regimes related to the flexibility parameter much greater than $0.241$.

The one-dimensional model developed by Christov {\it et. al.} \cite{christov_flow_2018} presents the concept of the flexibility parameter, giving a deep insight into the underlying physics.
It also provides a governing parameter that helps the engineers to design the deformable microchannels.  
The other advantage of the one-dimensional model 
in comparison with the three-dimensional coupled fluid-solid mechanics model 
is its convenience, since the one-dimensional model does not need the (several-hour) time-consuming simulations \cite{shidhore_static_2018}, yet it predicts the main fluid-structural characteristics acceptably with no fitting parameters.
On the other hand,
three dimensional model is a useful tool when the detailed information of the fluid-solid characteristics are desired, 
particularly when the effects of elastic membrane clamping at inlet and outlet, drag at the sidewalls of a microchannel that is not sufficiently shallow, etc., are needed to be taken into account.

In case that three-dimensional model is preferred, 
the threshold of $\chi\approx 0.241$ can be considered as a guideline.
According to Figs.~\ref{Fig-Straight_pressure_vfr-00} (a) and \ref{Fig-Res-lin-press}, the linear pressure variation assumption is reasonably acceptable when $\chi<0.241$,
practically decoupling 
the fluid and solid aspects of the problem.
The three-dimensional solid mechanics model, where a linear pressure variation is applied to the membrane, can then be used in order to obtain more accurate details of the structural characteristics. 
Similarly, the obtained deformed shape of the membrane 
can be used as the ceiling of a {rigid} microchannel for the three-dimensional fluid mechanics modeling if the details of the fluid flow field are needed.

When their requirements are met, such one-way-coupled strategies significantly reduce the computational costs, while they provide acceptably accurate details of the fluid-solid mechanics.
A similar approach was used by Ozsun~{\it et. al.} \cite{ozsun_non-invasive_2013}~in order to find the pressure distribution within deformable microchannels through two-dimensional fluid flow simulations. 
They replaced the ceiling with a rigid wall, but the deformed wall shape was preserved by importing the experimentally measured profile of the deformed membrane into the simulation.

\section[Conclusion]{Conclusion}
\label{Conclusion}

{Volumetric flow rate\textemdash pressure difference} relationship for the shallow deformable microchannels with ultra-low height-to-width-ratios was investigated in this work.
Our work enabled us to study the low-Reynolds-number, {\it i.e.} Re$\sim O(10^{-2}$\textemdash$10^{+1})$, fluid flow in the deformable microchannel under the new regimes of the flexibility parameter ($\chi$) with values a-few-orders-of-magnitude greater than those currently available in the literature.
The experimental results together with the scale analysis of 
the one-dimensional model developed by Christov {\it et.al.} \cite{christov_flow_2018} based on the {\it lubrication theory} and {\it thin-membrane-bending framework}, 
revealed the various distinct fluid-structural characteristic behaviors under the different flexibility parameter regimes. 
The difference between the characteristic behavior of fluid flow through a deformable microchannel and that related to its rigid counterpart was reflected into $Q^\ast$, which represents the ratio of the volumetric flow rate through the deformable channel to that through its theoretically rigid counterpart.
A master curve was obtained for the fluid flow through any arbitrary shallow and long deformable microchannel via plotting $Q^\ast$ versus $\chi$.
It was shown that depending on the flexibility parameter's magnitude, 
altering the pressure difference across the shallow deformable microchannel can activate/inactivate the various orders of the flexibility parameter.
The shallow deformable microchannels act similar to their corresponding rigid microchannels under the {sufficiently} small flexibility parameter. 
In this regime, the volumetric flow rate is proportional to $WH^3_0$, where $W$ and $H_0$ respectively refer to the microchannel width and original height.
On the other hand, sufficiently increasing the pressure difference across the microchannel can result in the transition to the higher-flexibility regimes.
Under the {sufficiently} large pressure differences across the deformable microchannel, the effects of the microchannel original height on the fluid-solid characteristics disappear.
In these situations, the volumetric flow rate becomes proportional to $W^{13}$.
Such a significantly distinct behavior together with the fact that the flexibility parameter is proportional to $W^4$ suggest that two nearly identical microchannels with the slightly different widths ($\sim 60\%$) can have two flexibility parameters one order of magnitude different from each other, which can result in the two distinct fluid-solid characteristic behaviors under the same pressure difference across the channels.
It was also found that the linear pressure assumption through the deformable microchannel is acceptable for the regimes with the flexibility parameter smaller than $0.241$ with less than $1\%$ error in the predicted volumetric flow rate, while this assumption is noticeably erroneous for the regimes with the large flexibility parameter.
For the cases that the flexibility parameter is smaller than $0.241$, the fluid and solid mechanics can be considered decoupled.

\section*{Acknowledgments}
\label{Acknowledgments}

Authors thank Dr. Baokang Bi and the staff of W. M. Keck Microfabrication Facility, and Karl Dersch and the staff of ECE Research Cleanroom in Michigan State University for their assistance with chemistry and cleanroom fabrication.

\bibliographystyle{spphys}       

\bibliography{Fluid_Flow_Shallow_Deformable_Channel}

\begin{thebibliography}{10}
\providecommand{\url}[1]{{#1}}
\providecommand{\urlprefix}{URL }
\expandafter\ifx\csname urlstyle\endcsname\relax
  \providecommand{\doi}[1]{DOI \discretionary{}{}{}#1}\else
  \providecommand{\doi}{DOI \discretionary{}{}{}\begingroup
  \urlstyle{rm}\Url}\fi

\bibitem{leslie_frequency-specific_2009}
D.C. Leslie, C.J. Easley, E.~Seker, J.M. Karlinsey, M.~Utz, M.R. Begley, J.P.
  Landers, Frequency-specific flow control in microfluidic circuits with
  passive elastomeric features, Nature Physics \textbf{5}(3), 231 (2009).
\newblock \doi{10.1038/nphys1196}.
\newblock \urlprefix\url{http://www.nature.com/doifinder/10.1038/nphys1196}

\bibitem{lam_microfluidic_2006}
E.W. Lam, G.A. Cooksey, B.A. Finlayson, A.~Folch, Microfluidic circuits with
  tunable flow resistances, Applied Physics Letters \textbf{89}(16), 164105
  (2006).
\newblock \doi{10.1063/1.2363931}.
\newblock \urlprefix\url{http://aip.scitation.org/doi/10.1063/1.2363931}

\bibitem{song_elastomer_2012}
W.~Song, A.E. Vasdekis, D.~Psaltis, Elastomer based tunable optofluidic
  devices, Lab on a Chip \textbf{12}(19), 3590 (2012).
\newblock \doi{10.1039/c2lc40481h}.
\newblock \urlprefix\url{http://xlink.rsc.org/?DOI=c2lc40481h}

\bibitem{abate_valve-based_2009}
A.R. Abate, M.B. Romanowsky, J.J. Agresti, D.A. Weitz, Valve-based flow
  focusing for drop formation, Applied Physics Letters \textbf{94}(2), 023503
  (2009).
\newblock \doi{10.1063/1.3067862}.
\newblock \urlprefix\url{http://aip.scitation.org/doi/10.1063/1.3067862}

\bibitem{raj_droplet_2016}
A.~Raj, R.~Halder, P.~Sajeesh, A.K. Sen, Droplet generation in a microchannel
  with a controllable deformable wall, Microfluidics and Nanofluidics
  \textbf{20}(7) (2016).
\newblock \doi{10.1007/s10404-016-1768-4}.
\newblock \urlprefix\url{http://link.springer.com/10.1007/s10404-016-1768-4}

\bibitem{huang_tunable_2009}
S.B. Huang, M.H. Wu, G.B. Lee, A tunable micro filter modulated by pneumatic
  pressure for cell separation, Sensors and Actuators B: Chemical
  \textbf{142}(1), 389 (2009).
\newblock \doi{10.1016/j.snb.2009.07.046}.
\newblock
  \urlprefix\url{http://linkinghub.elsevier.com/retrieve/pii/S092540050900611X}

\bibitem{gerhardt_chromatographic_2011}
T.~Gerhardt, S.~Woo, H.~Ma, Chromatographic behaviour of single cells in a
  microchannel with dynamic geometry, Lab on a Chip \textbf{11}(16), 2731
  (2011).
\newblock \doi{10.1039/c1lc20092e}.
\newblock \urlprefix\url{http://xlink.rsc.org/?DOI=c1lc20092e}

\bibitem{beattie_clog-free_2014}
W.~Beattie, X.~Qin, L.~Wang, H.~Ma, Clog-free cell filtration using resettable
  cell traps, Lab on a Chip \textbf{14}(15), 2657 (2014).
\newblock \doi{10.1039/c4lc00306c}.
\newblock \urlprefix\url{http://xlink.rsc.org/?DOI=c4lc00306c}

\bibitem{huang_clogging-free_2014}
S.B. Huang, Y.~Zhao, D.~Chen, H.C. Lee, Y.~Luo, T.K. Chiu, J.~Wang, J.~Chen,
  M.H. Wu, A clogging-free microfluidic platform with an incorporated
  pneumatically driven membrane-based active valve enabling specific membrane
  capacitance and cytoplasm conductivity characterization of single cells,
  Sensors and Actuators B: Chemical \textbf{190}, 928 (2014).
\newblock \doi{10.1016/j.snb.2013.09.070}.
\newblock
  \urlprefix\url{http://linkinghub.elsevier.com/retrieve/pii/S0925400513011118}

\bibitem{javanmard_targeted_2007}
M.~Javanmard, A.H. Talasaz, M.~Nemat-Gorgani, F.~Pease, M.~Ronaghi, R.W. Davis,
  Targeted cell detection based on microchannel gating, Biomicrofluidics
  \textbf{1}(4), 044103 (2007).
\newblock \doi{10.1063/1.2815760}.
\newblock \urlprefix\url{http://aip.scitation.org/doi/10.1063/1.2815760}

\bibitem{wang_ultrasensitive_2013}
T.~Wang, M.~Zhang, D.D. Dreher, Y.~Zeng, Ultrasensitive microfluidic
  solid-phase {ELISA} using an actuatable microwell-patterned {PDMS} chip, Lab
  on a Chip \textbf{13}(21), 4190 (2013).
\newblock \doi{10.1039/c3lc50783a}.
\newblock \urlprefix\url{http://xlink.rsc.org/?DOI=c3lc50783a}

\bibitem{smith_research_2014}
J.R. Smith, I.G. Arcibal, A.~Polini, M.R. Dokmeci, A.~Khademhosseini, Research
  highlights, Lab Chip \textbf{14}(1), 157 (2014).
\newblock \doi{10.1039/C3LC90120C}.
\newblock \urlprefix\url{http://xlink.rsc.org/?DOI=C3LC90120C}

\bibitem{jang_selective_2012}
K.~Jang, Y.~Tanaka, J.~Wakabayashi, R.~Ishii, K.~Sato, K.~Mawatari, M.~Nilsson,
  T.~Kitamori, Selective cell capture and analysis using shallow
  antibody-coated microchannels, Biomicrofluidics \textbf{6}(4), 044117 (2012).
\newblock \doi{10.1063/1.4771968}.
\newblock \urlprefix\url{http://aip.scitation.org/doi/10.1063/1.4771968}

\bibitem{cheung__2012}
P.~Cheung, K.~Toda-Peters, A.Q. Shen, \textit{{In}} situ pressure measurement
  within deformable rectangular polydimethylsiloxane microfluidic devices,
  Biomicrofluidics \textbf{6}(2), 026501 (2012).
\newblock \doi{10.1063/1.4720394}.
\newblock \urlprefix\url{http://aip.scitation.org/doi/10.1063/1.4720394}

\bibitem{ozsun_non-invasive_2013}
O.~Ozsun, V.~Yakhot, K.L. Ekinci, Non-invasive measurement of the pressure
  distribution in a deformable micro-channel, Journal of Fluid Mechanics
  \textbf{734} (2013).
\newblock \doi{10.1017/jfm.2013.474}.
\newblock
  \urlprefix\url{https://www.cambridge.org/core/journals/journal-of-fluid-mechanics/article/noninvasive-measurement-of-the-pressure-distribution-in-a-deformable-microchannel/CD698400F32F2530A984BA302A782699}

\bibitem{seker_nonlinear_2009}
E.~Seker, D.C. Leslie, H.~Haj-Hariri, J.P. Landers, M.~Utz, M.R. Begley,
  Nonlinear pressure-flow relationships for passive microfluidic valves, Lab on
  a Chip \textbf{9}(18), 2691 (2009).
\newblock \doi{10.1039/b903960k}.
\newblock \urlprefix\url{http://xlink.rsc.org/?DOI=b903960k}

\bibitem{hardy_deformation_2009}
B.S. Hardy, K.~Uechi, J.~Zhen, H.~Pirouz~Kavehpour, The deformation of flexible
  {PDMS} microchannels under a pressure driven flow, Lab Chip \textbf{9}(7),
  935 (2009).
\newblock \doi{10.1039/B813061B}.
\newblock \urlprefix\url{http://xlink.rsc.org/?DOI=B813061B}

\bibitem{kang_pressure-driven_2014}
C.~Kang, C.~Roh, R.A. Overfelt, Pressure-driven deformation with soft
  polydimethylsiloxane ({PDMS}) by a regular syringe pump: challenge to the
  classical fluid dynamics by comparison of experimental and theoretical
  results, RSC Advances \textbf{4}(7), 3102 (2014)

\bibitem{neelamegam_experimental_2015}
R.~Neelamegam, V.~Shankar, Experimental study of the instability of laminar
  flow in a tube with deformable walls, Physics of Fluids \textbf{27}(2),
  024102 (2015).
\newblock \doi{10.1063/1.4907246}.
\newblock \urlprefix\url{http://aip.scitation.org/doi/10.1063/1.4907246}

\bibitem{raj_hydrodynamics_2017}
M.K. Raj, S.~DasGupta, S.~Chakraborty, Hydrodynamics in deformable
  microchannels, Microfluidics and Nanofluidics \textbf{21}(4), 70 (2017).
\newblock \doi{10.1007/s10404-017-1908-5}.
\newblock
  \urlprefix\url{https://link.springer.com/article/10.1007/s10404-017-1908-5}

\bibitem{chakraborty_fluid-structure_2012}
D.~Chakraborty, J.R. Prakash, J.~Friend, L.~Yeo, Fluid-structure interaction in
  deformable microchannels, Physics of Fluids \textbf{24}(10), 102002 (2012).
\newblock \doi{10.1063/1.4759493}.
\newblock \urlprefix\url{http://aip.scitation.org/doi/10.1063/1.4759493}

\bibitem{shidhore_static_2018}
T.C. Shidhore, I.C. Christov, Static response of deformable microchannels: a
  comparative modelling study, Journal of Physics: Condensed Matter
  \textbf{30}(5), 054002 (2018).
\newblock \urlprefix\url{http://stacks.iop.org/0953-8984/30/i=5/a=054002}

\bibitem{gervais_flow-induced_2006}
T.~Gervais, J.~El-Ali, A.~Günther, K.F. Jensen, Flow-induced deformation of
  shallow microfluidic channels, Lab on a Chip \textbf{6}(4), 500 (2006).
\newblock \doi{10.1039/b513524a}.
\newblock \urlprefix\url{http://xlink.rsc.org/?DOI=b513524a}

\bibitem{christov_flow_2018}
I.C. Christov, V.~Cognet, T.C. Shidhore, H.A. Stone, Flow rate–pressure drop
  relation for deformable shallow microfluidic channels, Journal of Fluid
  Mechanics \textbf{841}, 267 (2018).
\newblock \doi{10.1017/jfm.2018.30}.
\newblock
  \urlprefix\url{https://www.cambridge.org/core/journals/journal-of-fluid-mechanics/article/flow-ratepressure-drop-relation-for-deformable-shallow-microfluidic-channels/798E21E2643E415BFA721DDD6298D4A6}

\bibitem{raj_flow-induced_2016}
A.~Raj, A.K. Sen, Flow-induced deformation of compliant microchannels and its
  effect on pressure–flow characteristics, Microfluidics and Nanofluidics
  \textbf{20}(2) (2016).
\newblock \doi{10.1007/s10404-016-1702-9}.
\newblock \urlprefix\url{http://link.springer.com/10.1007/s10404-016-1702-9}

\bibitem{mehboudi_two-step_2018}
A.~Mehboudi, J.~Yeom, A two-step sealing-and-reinforcement {SU}8 bonding
  paradigm for the fabrication of shallow microchannels, Journal of
  Micromechanics and Microengineering \textbf{28}(3), 035002 (2018).
\newblock \urlprefix\url{http://stacks.iop.org/0960-1317/28/i=3/a=035002}

\bibitem{gere_mechanics_2004}
J.M. Gere, \emph{Mechanics of materials}, 6th edn. (Brooks/Cole-Thomas
  Learning, Belmont, CA, 2004)

\bibitem{luo_simple_2003}
C.~Luo, T.W. Schneider, R.C. White, J.~Currie, M.~Paranjape, A simple
  deflection-testing method to determine {Poisson} s ratio for {MEMS}
  applications, Journal of Micromechanics and Microengineering \textbf{13}(1),
  129 (2003).
\newblock \doi{10.1088/0960-1317/13/1/318}.
\newblock
  \urlprefix\url{http://stacks.iop.org/0960-1317/13/i=1/a=318?key=crossref.81166d0e06484fabd83d63f992861edc}

\end{thebibliography}

\clearpage

\clearpage

\clearpage

\clearpage

\clearpage

\clearpage

\clearpage

\clearpage

\clearpage

\end{document}